\documentclass[aps,amssymb,amsmath,raggedbottom,reprint,superscriptaddress]{revtex4-1}

\usepackage{graphicx}
\usepackage{epstopdf}
\usepackage{braket}
\usepackage{color}

\newcommand{\K}{\mathcal K}

\newcommand{\V}{\mathcal V}
\newcommand{\G}{\mathcal G}

\newcommand{\R}{\mathcal R}
\newcommand{\E}{\mathcal E}
\newcommand{\OO}{\mathcal O}
\newcommand{\LL}{\mathcal L}

\newcommand{\rr}{\mathbf r}

\newcommand{\clr}[1]{{\color{black} #1}}

\newcommand{\PP}{P}

\newcommand{\rmd}{\,{\rm d}}
\newcommand{\bp}{\overline p}

\newcommand{\vol}{\operatorname{vol}}

\begin{document}


\title{Entropy of Spatial Network Ensembles}


\author{Justin P. Coon}
\affiliation{Department of Engineering Science, University of Oxford, Parks Road, Oxford OX1 3PJ, UK}

\author{Carl P. Dettmann}
\affiliation{School of Mathematics, University of Bristol, University Walk, Bristol BS8 1TW, UK}

\author{Orestis Georgiou}
\affiliation{School of Mathematics, University of Bristol, University Walk, Bristol BS8 1TW, UK}
\affiliation{Ultrahaptics, The West Wing, Glass Wharf, Bristol BS2 0EL, UK}


\date{\today}

\begin{abstract}
We analyze complexity in spatial network ensembles through the lens of graph entropy.  Mathematically, we model a spatial network as a \clr{\emph{soft} random geometric graph, i.e., a graph with two sources of randomness, namely nodes located randomly in space and links formed independently between pairs of nodes with probability given by a specified function (the ``pair connection function'') of their mutual distance}.  We consider the general case where randomness arises in node positions as well as pairwise connections (i.e., for a given pair distance, the corresponding edge state is a random variable).  Classical random geometric graph and exponential graph models can be recovered in certain limits.  We derive a simple bound for the entropy of a spatial network ensemble and calculate the conditional entropy of an ensemble given the node location distribution for hard and soft (probabilistic) pair connection functions.  Under this formalism, we derive the connection function that yields maximum entropy under general constraints.  Finally, we apply our analytical framework to study two practical examples: ad hoc wireless networks and the US flight network.  Through the study of these examples, we illustrate that both exhibit properties that are indicative of nearly maximally entropic ensembles.
\end{abstract}

\pacs{???}

\maketitle

\section{\label{sec:i}Introduction}
The topological structure of networks has been studied for many years through the lens of graph entropy~\cite{Simonyi1995}.  This formalism, which is deeply rooted in statistical physics and information theory, allows one to quantitatively characterize the complexity or inherent information content of systems that can be described by a graphical model~\cite{Mowshowitz1968a,Mowshowitz1968b,Mowshowitz1968c,Mowshowitz1968d,Anand2009,Anand2011}.  Applications of entropy-based methods to the study of networked systems are abundant and include problems related to molecular structure classification~\cite{Bonchev1983}, social networks~\cite{Everett1985,Dehmer2011}, data compression~\cite{Choi2012}, quantum entanglement~\cite{deBeaudrap2016,Simmons2017}, and topological uncertainty in communication networks~\cite{Coon2016,Coon2017a,Coon2017b}.   Of course, this diverse range of applications has led to the definition of numerous entropic measures~\cite{Dehmer2008}, and the study and unification of such measures continues to garner interest~\cite{Fehr2014}.

To date, research on graph entropy has been mostly focused on systems that do not depend on an underlying spatial embedding, or for which this embedding has been abstracted or ignored.  In this context, numerous measures of structural entropy have been developed that allow one to quantify the entropy of a single graph~\cite{Dehmer2008,Dehmer2011}.  This is typically done by identifying some characteristic of the network that is of interest, and defining a probability distribution on this characteristic by using a frequency interpretation of occurring events.  An alternative approach to studying graph entropy is to consider a measure on the entire ensemble, not just a single graph.  In this case, a probability distribution is defined on the ensemble, and various measures of entropy (e.g., Gibbs entropy, Shannon entropy, von Neumann entropy, R\'enyi entropy) can be calculated for the ensemble~\cite{Anand2009}.  The canonical example that one might begin with is the Erd\H{o}s-R\'enyi (ER) model, for which each edge exists independently with probability $p$.  It is straightforward to observe that the probability of a particular graph $G$ containing $n$ nodes and $k$ edges is just
\begin{equation}
  P(G) = p^k (1-p)^{\clr{c_n}-k}
\end{equation}
\clr{where $c_n = n(n-1)/2$}.  Hence, a probability distribution is well defined on the graph ensemble $\G$ for $k=0,\ldots,\clr{c_n}$ since $P(G) \geq 0$ for all $G\in\G$ and $\sum_{G\in\G}P(G) = 1$. Suppose we wish to calculate the Shannon entropy of the ensemble $\G$, which admits the expression
\begin{equation}\label{eq:H(G)_1}
  H(\G) = -\sum_{G\in \G} \PP(G) \ln \PP(G).
\end{equation}
By enumerating the different combinations of edges that form the graphs in $\G$, it follows that the Shannon entropy of the ER ensemble can be written as
\begin{equation}\label{eq:H(G)_2}
  H(\G) =  -c_n (p\ln p + (1-p)\ln(1-p)).
\end{equation}

In contrast to the spatially independent models discussed above, geography has been shown to play an important role in many engineered, physical and social networks~\cite{Barthelemy2011}.  For example, an empirical study conducted in Belgium and reported in~\cite{Lambiotte2008} demonstrated that social connectivity between two individuals decays like $r^{-2}$ over distances from about 1 km to 100 km.  Spatial network models have also been employed to study the spread of epidemics~\cite{Riley2007}, mobile phone viruses~\cite{Wang2009}, and to analyze connectivity in wireless communication networks~\cite{Coon2012a}.  Local structural observables that describe these networks, such as the clustering coefficient and the degree distribution, are well understood for simple models.  However, while some work has been done to characterize the entropy of spatial networks~\cite{Graff2008}, relatively little research has focused on the entropy of spatial network ensembles.  The notable exception to this lack of available results in the literature is~\cite{Halu2014} and the works pertaining to wireless communication networks by the first author~\cite{Coon2016,Coon2017a,Coon2017b}.

In this contribution, we characterize the Shannon entropy of spatial network ensembles by employing a \emph{soft} random geometric graph (RGG) framework~\clr{\cite{Penrose2016}}.  Randomness in this context is derived both from the node positions in space as well as from inherent uncertainties associated with the edge states.  We begin by providing an upper bound on spatial network ensemble entropy, which is derived from a simple result from information theory.  \clr{We then discuss entropy in the \emph{conditional} sense, whereby we assume some prior knowledge of the underlying node locations and average the entropy over the spatial distribution.}  In many physical systems, such knowledge is available; hence, this measure of conditional entropy lends itself nicely to practical interpretations. 

\clr{Our approach to characterizing network entropy takes a different route than the work reported in~\cite{Halu2014}.  Specifically,~\cite{Halu2014} considered the case where node positions are static and known \emph{a priori}, which enables the study of systems with link specific constraints and even ensembles of multiplexes.  In contrast, we focus on extracting information about systems for which node locations are, themselves, a source of randomness by averaging over all possible spatial node configurations.  This is a conventional information theoretic approach to understanding how one random variable yields information about another; hence, our work can be viewed as a step towards a larger information theoretic framework for spatial networks.}  Indeed, using our formalism facilitates the calculation of the mutual information (more generally, the relative entropy) between the network topology and the underlying point process that statistically characterizes node locations~\cite{Cover2012}.  Under this formalism, we derive the connection function that yields maximum entropy under general constraints.  We then apply our analytical methods to study two practical examples: ad hoc wireless networks and the conterminous US flight network.  We illustrate the intriguing result that both systems exhibit properties that are indicative of nearly maximally entropic ensembles.

\section{\label{sec:e}Entropy}
Consider a set $\V$ of $n$ nodes embedded in a space $\K\subset\mathbb R^d$, which has volume $K = \vol(\K)$.  The locations of the nodes are denoted by $\rr_1,\ldots,\rr_n$, and these locations form a point process in $\K$.  In this work, we will consider a simple, uniform point process, which implies the spatial distribution of nodes can be described by a constant, finite intensity of $\rho = n/K$ nodes per unit volume.  In what follows, it will suffice to assume that $n$ is fixed.  Hence, the triple $(\V,\K,\rho)$ describes a binomial point process. Similar results to those disclosed herein follow when $n$ is Poisson distributed.  

An (undirected) edge exists between nodes $i$ and $j$ in $\V$ with probability $p(r_{i,j})$, where $r_{i,j}$ denotes the distance between points $\rr_i$ and $\rr_j$.  In this work, we assume the spatial embedding is Euclidean, i.e., $r_{i,j} = \Vert\rr_i - \rr_j\Vert$.  \clr{In the case where $p(r_{i,j})$ is an indicator function that is one for $r < r_0$ and zero otherwise, with $r_0$ denoting the maximum connection range, we recover the hard disk model used in the classical RGG formalism~\cite{Penrose2003}.  For other connection functions $p$ -- e.g., monotonically decreasing functions in the distance argument -- we have the so-called \emph{soft} RGG model~\cite{Penrose2016,Dettmann2016}.}  As discussed above, we denote the set of all graphs by $\G$.  For a particular graph $G\in\G$, the set of edges is signified by $\E_G$.  Note that these sets are defined without reference to a particular underlying spatial embedding, or even the spatial distribution of the vertices.

We are interested in quantifying the Shannon entropy of the ensemble $\G$.  Each graph $G\in\G$ occurs with probability $\PP(G)$, which depends upon the spatial distribution of the nodes and the pair connection function $p$.  The definition of the Shannon entropy of $\G$ was given in eq.~(\ref{eq:H(G)_1}).  In contrast to the example of ER graph ensembles, the existence of edge $(i,j)$ is a function of the pair distance $r_{i,j}$.    Viewing a graph as a random variable $G$ with support $\G$, we can easily deduce that the distribution of $G$ is equivalent to the distribution of the edge set only, since the spatial embedding of the vertices is captured in the edge probabilities.  Let $X_{i,j}$ denote a Bernoulli random variable  that models the existence of edge $(i,j)$.  It follows that 
we can write the entropy of the network ensemble as the joint entropy of the sequence $\{X_{i,j}\}$, i.e.,
\begin{equation}
  H(\G) = H(X_{1,2},X_{1,3},\ldots,X_{n-1,n}).
\end{equation}
We can now invoke the well known independence bound on joint Shannon entropy to obtain the inequality
\begin{equation}\label{eq:H(G)_3}
  H(\G) \leq \sum_{i<j} H(X_{i,j})
\end{equation}
where equality holds if all $\{X_{i,j}\}$ are independent.  A more thorough treatment of this bound in the context of spatial network entropy is provided in~\cite{Coon2016}.

The random variable $X_{i,j}$ is physically related to nodes $i$ and $j$, but it should be stressed that $\PP(X_{i,j} = 1)\neq p(r_{i,j})$, since pair distance information is marginalized in eq.~(\ref{eq:H(G)_3}).  More accurately, we can write
\begin{equation}\label{eq:p(r)}
  \PP(X_{i,j} = 1) = \int_0^D f(r_{i,j}) p(r_{i,j}) \rmd r_{i,j}
\end{equation}
which is just the probability that edge $(i,j)$ exists averaged over all pair distances, where $D = \sup_{\rr_i,\rr_j\in\K}\Vert \rr_i - \rr_j \Vert$ is the diameter of the domain $\K$ and $f(r_{i,j})$ is the pair distance probability density corresponding to nodes $i$ and $j$.  \clr{By noting that} the pair distance density and the pair connection function are, respectively, identical for all $i\neq j$, we can simply let $\bp := \PP(X_{i,j} = 1)$ and write
\begin{equation}\label{eq:HGbound}
  H(\G) \leq c_n H_2(\bp)
\end{equation}
where
\begin{equation}
  H_2(\bp) = -\bp \ln\bp - (1-\bp)\ln(1-\bp)
\end{equation}
is the binary entropy function.  So by assuming pair distances are independent in the \clr{soft RGG} case, we obtain an ER-like result, but where the pair connection probability is averaged over the pair distance distribution.  Clearly, when $\bp = 1/2$, the bound is maximized.


\section{\label{sec:ce}Conditional Entropy}
To enable us to study the effect of a particular embedding on the entropy of \clr{a soft RGG} ensemble, we turn to the information theoretic notion of \emph{conditional entropy}.  In the context of our problem, the conditional entropy of the graph ensemble given the distribution of vertex locations is defined as
\begin{equation}
  H(\G|\R) =  \braket{H(\G|\rr_1,\ldots,\rr_n)} 
\end{equation}
where the notation $\braket{\cdot}$ denotes the average of a function over the vertex positions, which for uniformly distributed vertices in $\K$ is given by
\begin{equation}
  \braket{F(\rr_1,\ldots,\rr_n)} = \frac{1}{K^n}\int_{\K^n} F(\rr_1,\ldots,\rr_n) \rmd \rr_1\cdots \rmd \rr_n.
\end{equation}
This quantity can be more conveniently expressed in terms of an average over the pair distances
\begin{equation}
  H(\G|\R) =  \braket{H(\G|r_{1,2},r_{1,3},\ldots,r_{n-1,n})}
\end{equation}
where the average is defined in the appropriate manner with respect to the joint pair distance density $f(r_{1,2},r_{1,3},\ldots,r_{n-1,n})$.  By using a similar argument to that employed in eq.~(\ref{eq:H(G)_3}), it is possible to show that
\begin{equation}\label{eq:sum_HXr}
  H(\G|r_{1,2},\ldots,r_{n-1,n}) = \sum_{i<j}H(X_{i,j}|r_{i,j})
\end{equation}
where equality follows since, in this case, the edge states $\{X_{i,j}\}$ are independent conditioned on the pair distances $\{r_{i,j}\}$.  Averaging the right-hand side of eq.~(\ref{eq:sum_HXr}) over the density $f(r_{1,2},r_{1,3},\ldots,r_{n-1,n})$ leads naturally to
\begin{equation}\label{eq:HGR}
  H(\G|\R) =  c_n \int_0^D f(r) H_2(p(r)) \rmd r.
\end{equation}

A fundamental result of information theory states that conditioning reduces uncertainty~\clr{\cite{Cover2012}}.  \clr{Combining this notion with eq.~\eqref{eq:HGbound} manifests in the relation}
\begin{equation}\label{eq:inequalities}
  H(\G|\R) \leq H(\G) \leq c_n  H_2(\bp).
\end{equation}
In fact, \clr{the lower and upper bounds} follow directly from the concavity of Shannon entropy and Jensen's inequality, i.e., the left-hand side is the average of the entropy, whereas the right-hand side is the entropy of the average.  The tightness of each bound depends upon the pair connection function $p(r)$.  In practical networks, such as social networks or communication networks, the underlying system parameters dictate the form of this function.  In general, very soft pair connection functions lead to relatively tight bounds in this context since there is little dependence upon the spatial embedding.  On the other hand, if \clr{$p(r) = 1$ for $0\leq r < r_0$ and $p(r) = 0$ otherwise, then we recover the classical RGG formalism and $H(\G|\R) = 0$, but the upper bound on $H(\G)$ is still $\OO(n^2)$}.

\section{\label{sec:maxent}Entropy Maximizing Connection Functions}
Understanding how different parameters affect uncertainty in spatial networks is of great importance.  Hence, it is of interest to determine the function $p(r)$ that maximizes the conditional entropy of the network ensemble.  Without further information, we can immediately observe that the entropy maximizing function (in the conditional sense) is just $p(r) = 1/2$, in which case $H(\G|\R) = H(\G) = c_n \ln 2$.  This is rather uninteresting, however, since we are back to the ER graph ensemble in which the spatial embedding is irrelevant.  To make progress beyond this benchmark, we note that practical spatial networks typically exhibit certain properties, which we may wish to incorporate into the maximization task through constraints.  For example, the mean degree of a geometric graph ensemble can be written as
\begin{equation}
  \bar\delta = (n-1) \int_0^D f(r) p(r)\rmd r.
\end{equation}
Considering a network with a given mean degree, we can formulate a constrained variational problem, which has Lagrangian
\begin{equation}
  \LL(p) = c_n f(r) H_2(p(r)) - \mu (n-1) f(r) p(r)
\end{equation}
with $\mu$ denoting the required multiplier for the mean degree constraint.  Setting $\LL_p = 0$ and using the constraint to solve for the multiplier yields the stationary function $p = \bar\delta/(n-1)$ for all $r\in [0,D]$, and thus we again see that the spatial embedding of the graph does not affect the maximum entropy probability.  Hence, \clr{by using eqs.~(\ref{eq:p(r)}), (\ref{eq:HGbound}), and (\ref{eq:HGR}), we arrive at the} entropy relation
\begin{equation}
  H(\G|\R) = H(\G) \sim \frac{\bar\delta}{2}n\ln n
\end{equation}
for large $n$.  Although the previous two examples are independent of the spatial embedding and are, in some sense, uninteresting in the context of the present discussion, it is worth noting the difference in entropies for the dense and the sparse cases: $H(\G)=\OO(n^2)$ for the dense network with $p = 1/2$ and $H(\G)=\OO(n\ln n)$ for the sparse network with $p=\OO(n^{-1})$.

Many geometric networks observed in our physical world are characterized by more than just the mean degree.  Other local features often play an important role in the overall structure of the network.  For example, in wireless communication networks, pairwise connections often have a well-defined statistical make-up, which is governed by the modulation/demodulation techniques used at each node and the arrangement of scatterers in the environment~\cite{Coon2012}.  From a more general perspective, we may consider a graph ensemble where the pair connection function obeys a set of constraints
\begin{equation}\label{eq:constraints}
  \int_0^D \theta_\ell(r) p(r) \rmd r = k_\ell,\quad \ell = 1,\ldots,L
\end{equation}
where $\{\theta_\ell\}$ and $\{k_\ell\}$ are independent of $n$.  In such a case, we might seek the maximizing function $p(r)$ subject to these constraints.  We now have a constrained variational problem comprised of eq.~(\ref{eq:HGR}) and the constraints given in eq.~(\ref{eq:constraints}).  Solving the associated Euler-Lagrange equation yields the stationary function
\begin{equation}\label{eq:ferm_p}
  p(r) = \frac{1}{e^{\psi(r)} + 1}
\end{equation}
where
\begin{equation}
  \psi(r) = \frac{1}{c_n f(r)}\sum_{\ell=1}^L \lambda_\ell \theta_\ell(r)
\end{equation}
and $\{\lambda_\ell\}$ are the undetermined multipliers.  This solution points to explicit dependence on the spatial embedding through the pair distance density $f$ and the constraints captured by $\{\theta_\ell\}$.  \clr{Note that if $\theta_\ell(r) = g_\ell(r) f(r)$ for some function $g_\ell$ and $\ell = 1,\ldots,L$, the maximizing pair connection function is independent of $f$.  Of course, the multipliers $\{\lambda_\ell\}$ still depend upon the geometric properties of the network in this case.}  It will have occurred to many readers that eq.~(\ref{eq:ferm_p}) resembles the classical entropy maximizing distribution pertaining to quantum state occupancy in systems of noninteracting fermions.  Park and Newman pointed out a similar relation for exponential random graphs in~\cite{Park2004}.  Here, we have arrived at a similar result for spatial network ensembles in bounded domains.

For finite network domains, $f$ is independent of $n$.  Assuming for large $n$ that $\psi =\OO(n^{-\epsilon})$ for $\epsilon > 0$, we have $p(r) = 1/2 + \OO(n^{-\epsilon})$.  But this yields a degenerate outcome in which the constraints are independent of the pair connection function.  Hence, we must have that $\psi(r)=\OO(1)$, i.e., the multipliers scale like $n^2$.  The entropy maximizing connection function yields a dense network in the large $n$ limit, i.e., $H(\G)=\OO(n^2)$. The mean degree of this network is asymptotically
\begin{equation}
  \bar\delta \sim n \int_0^D \frac{f(r)}{e^{\psi(r)}+1}\rmd r.
\end{equation}

In finite networks, it is meaningful to consider adding a mean degree constraint to the set of constraints defined in eq.~(\ref{eq:constraints}).  In this case, the entropy maximizing connection function is
\begin{equation}
  p(r) = \frac{1}{e^{\psi(r) + {2\mu}/{n}} + 1}.
\end{equation}
Applying a mean degree constraint induces sparsity if the constraint is kept fixed for any $n$.  However, positive constraints cannot be satisfied in the large $n$ limit for $p\sim 1/n$ (see eq.~(\ref{eq:constraints})).  Rigorously, this follows from a fundamental result from calculus, which states that for all smooth functions $\xi(r)$ on the interval $[0,D]$, $\int_0^D \xi(r) p(r)\rmd r = 0$ if and only if $p(r) = 0$ on the interval.  Hence, stationary functions do not exist for some constraint conditions.

\section{\label{sec:ex}Examples}
Various examples of real-world spatial networks exist, and each is governed by particular connection rules and design constraints.  Here, we explore two such examples.

\subsection{\label{sec:ex1}Wireless Communication Networks}
The first example we will explore is related to ad hoc deployments of electronic devices that have the capability to communicate wirelessly.  In this example, a pairwise connection between two nodes separated by a distance $r$ forms with probability~\cite{Coon2012a}
\begin{equation}\label{eq:pr_rayleigh}
  p(r) = e^{-(r/r_0)^\eta}.
\end{equation}
The parameter $r_0$ denotes the typical connection range; it depends on physical quantities such as the wavelength of the transmission and antenna gains at the transmitter and receiver.  The parameter $\eta >0$, known as the \emph{path loss exponent}, is typically an experimentally determined number that indicates how quickly a transmission is attenuated as it propagates through the wireless medium.  Eq.~(\ref{eq:pr_rayleigh}) follows from information theoretic arguments, but possesses two convenient properties: (1) it is intuitive in that it exhibits a connection rule that captures a monotonically decreasing probability of pairwise connectivity as the distance between nodes increases; and (2) the parameter $\eta$ serves to define the degree of pairwise uncertainty experienced in the network, with $\eta\to\infty$ signifying a hard connection rule whereby two nodes are connected with probability one if they are separated by less than $r_0$ and are not connected otherwise, i.e., $\eta\to\infty$ signifies the \clr{classical RGG formalism}.

\begin{figure}[t!]
\includegraphics[width=8cm]{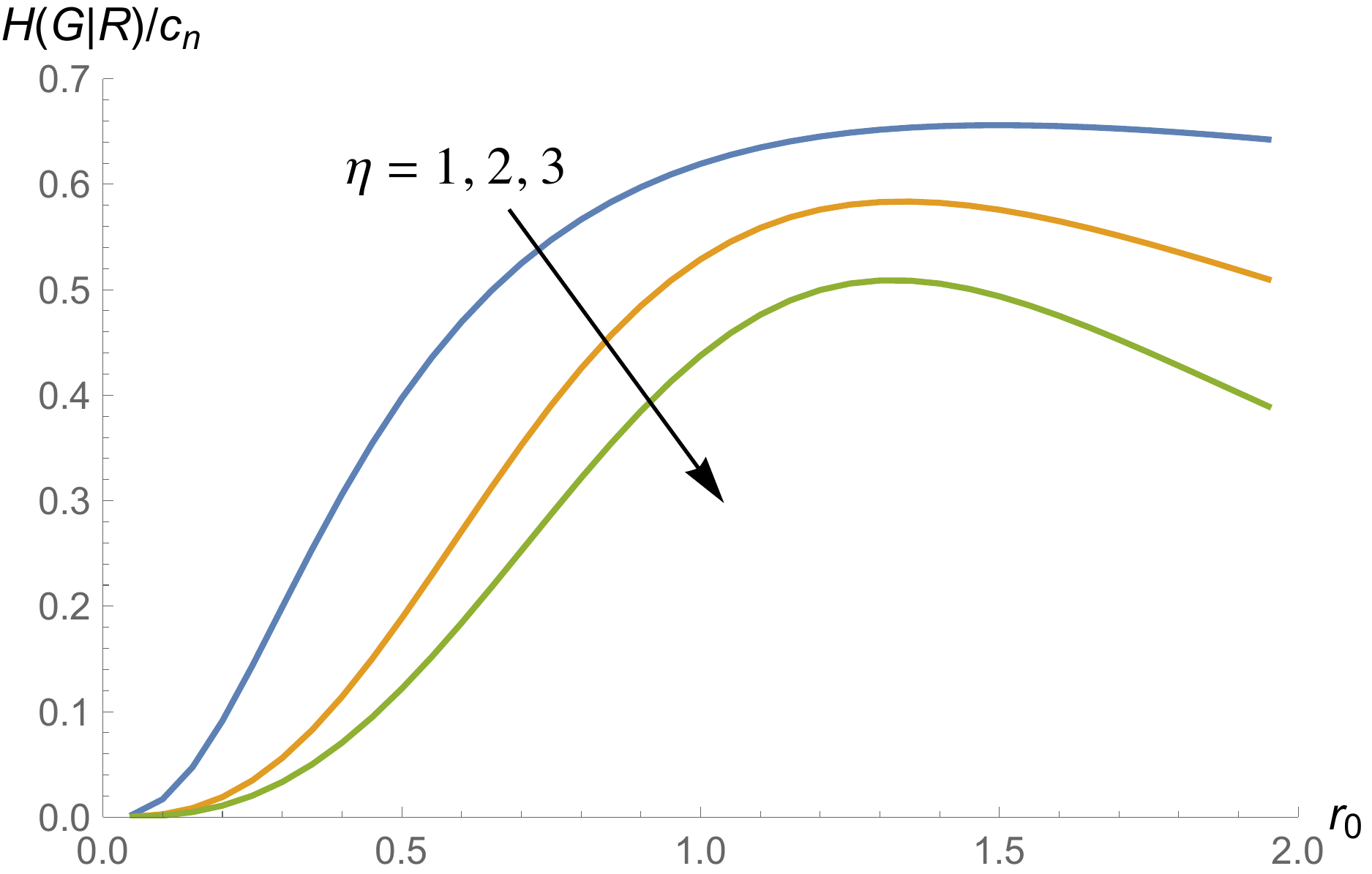}%
\caption{\label{fig:wireless1}Conditional entropy normalized with respect to the total number of possible edges plotted against the typical connection range in a wireless communication network.  The network domain is a sphere of unit radius in three dimensions; $f(r) = 3 r^2 (1 - 3 r/4 + r^3/16)$.  Three different exponents $\eta = 1,2,3$ were used to show how an increase in edge certainty (increasing $\eta$) reduces entropy.}
\end{figure}

Fig.~\ref{fig:wireless1} illustrates the effect that certainty in each pairwise connection has on the conditional entropy.  This figure relates to a network residing in a sphere of unit radius in three dimensions.  As we increase $\eta$, the pairwise uncertainty decreases and the conditional entropy of the ensemble correspondingly decreases.  Clearly, as $\eta\to\infty$, the conditional entropy tends to zero, although the unconditional entropy $H(\G)$ remains positive.

\begin{figure}[t!]
\includegraphics[width=8cm]{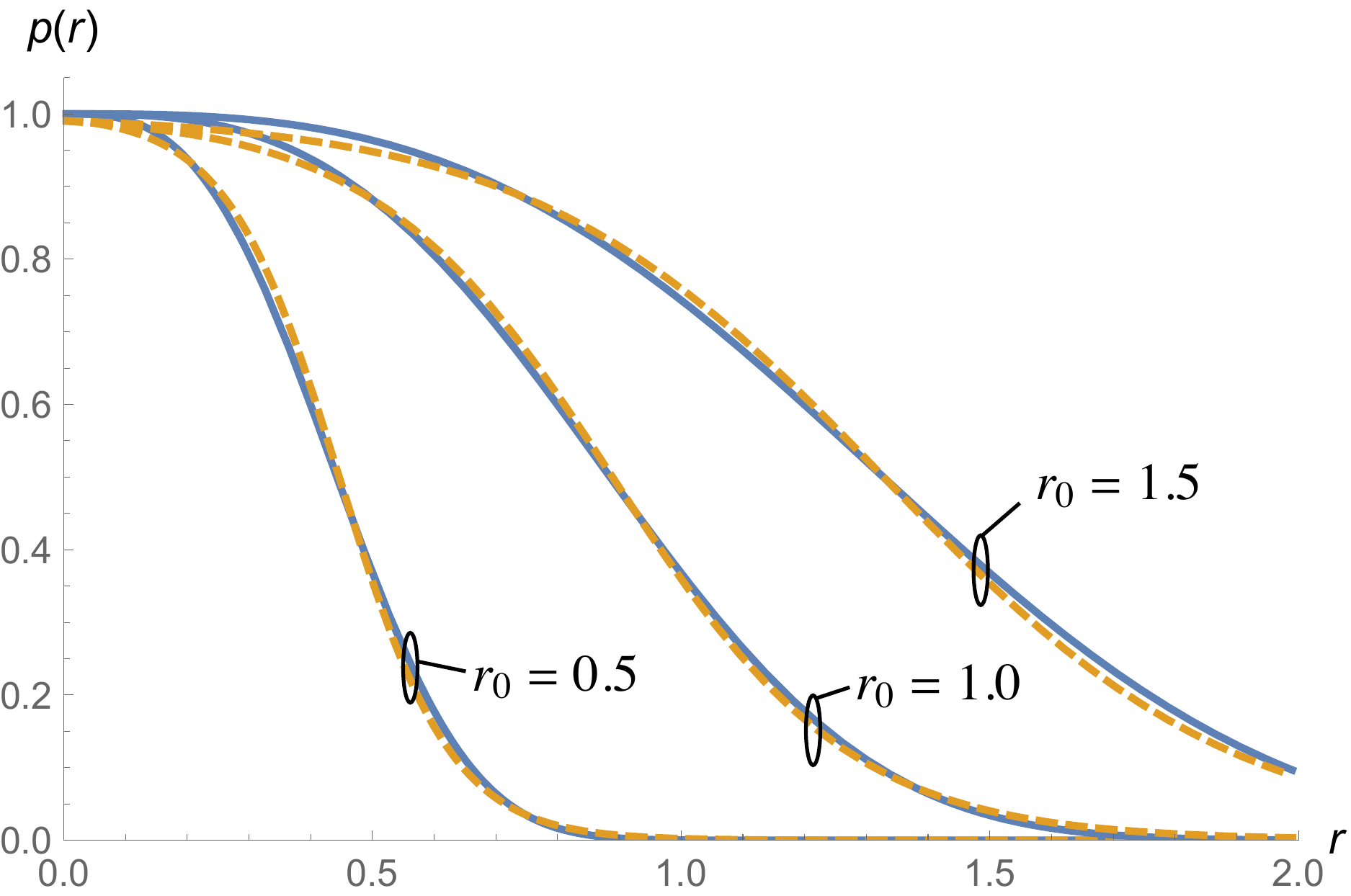}%
\caption{\label{fig:wireless2}Test function defined in eq.~(\ref{eq:pr_rayleigh}) (solid lines) and maximum conditional entropy function (dashed lines) plotted as a function of node separation for a network of $n = 10$ nodes.  For the test function, $\eta = 3$ and $r_0 = 0.5,\,1.0,\,1.5$.  Average connection probability and correlation constraints were imposed for the maximization.}
\end{figure}

In wireless networks, it is natural to consider a system in which both the average pair connection function and the correlation between the connection function and distance are constrained, i.e.,
\begin{equation}\label{eq:wireless_constraints}
  k_\ell = \int_0^D r^{\ell - 1}p(r) f(r) \rmd r,\quad \ell = 1,2.
\end{equation}
The correlation constraint somewhat captures the monotonically decreasing nature of the likelihood of pairwise connectivity as node separation increases, whereas the average connection probability constraint gives an indication of average performance in the network.  Let us consider the exponential test function given in eq.~(\ref{eq:pr_rayleigh}) with $\eta = 3$.  For each value of $r_0$, one can easily calculate the integrals in eq.~(\ref{eq:wireless_constraints}). Here, we consider the case where the network domain is a sphere of unit radius in three dimensions; $f(r) = 3 r^2 (1 - 3 r/4 + r^3/16)$.  This sets the constraints, which can be used to solve for the undetermined multipliers that define the entropy maximizing function (eq.~(\ref{eq:ferm_p})).  In this way, we can compare the test function and the maximum entropy connection probability as a function of distance $r$ (Fig.~\ref{fig:wireless2}).  Our analysis provides an interesting approach to studying complexity in wireless networks by allowing one to compare practical connection functions to the maximum entropy curves.  In the example discussed here, it is clear that the practical test function is extremely close to the entropy maximizing function.  In fact, the test functions depicted in Fig.~\ref{fig:wireless2} yield conditional entropies of over $99.4\%$ of the maximum in each case.  This result points to the intriguing conclusion that wireless networks adhering to the model studied here are nearly maximally complex \clr{(under the specified constraints)}.

\subsection{\label{sec:ex2}Flights in the United States}
\begin{figure}
\includegraphics[width=8cm]{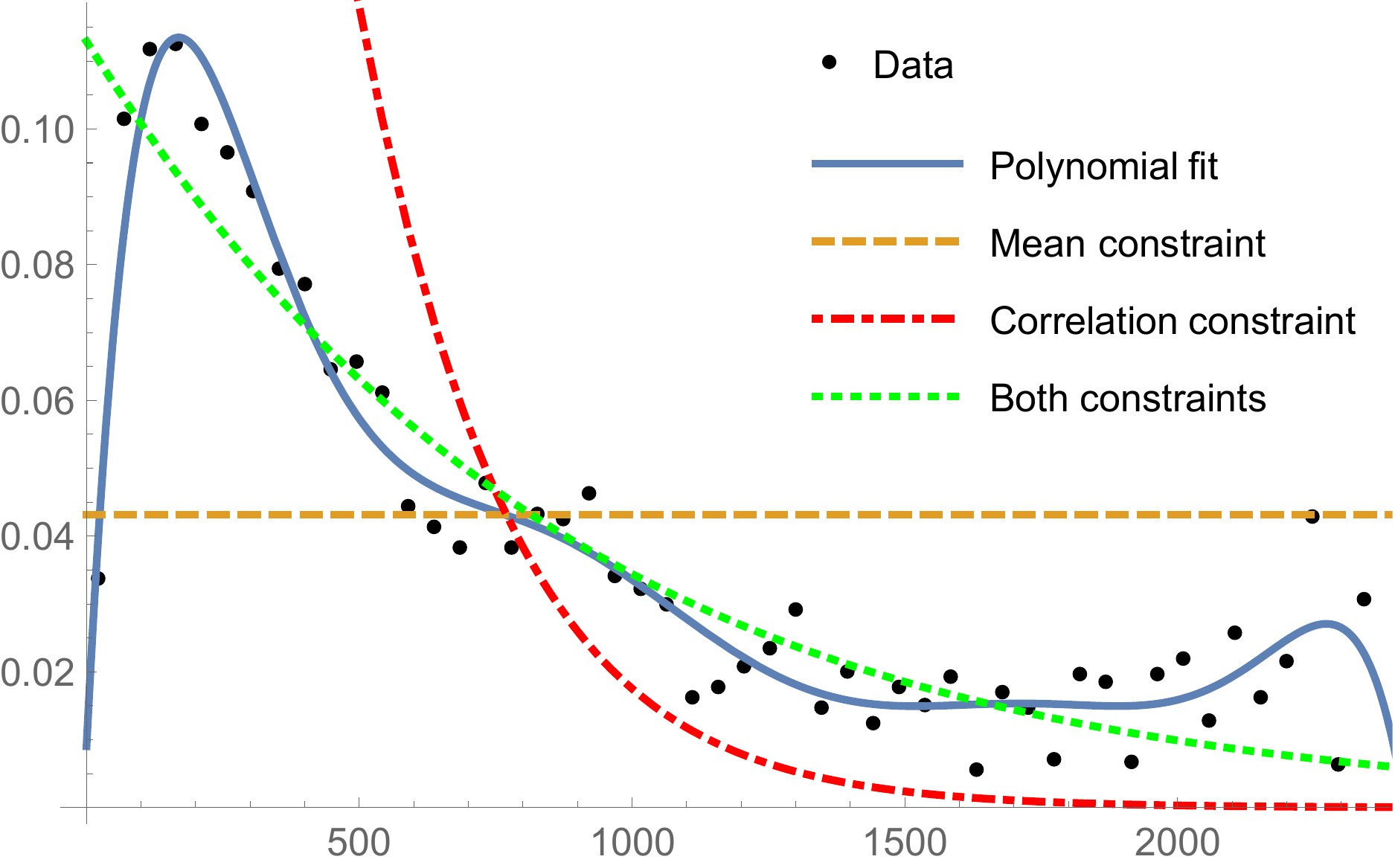}%
\caption{\label{fig:flights}Pair connection function derived from US flight data (solid line) and maximum conditional entropy functions \clr{for different sets of constraints} (dashed lines) plotted as a function of the distance between primary airports in the conterminous US.  \clr{Markers indicate raw histogram data obtained from~\cite{OFAirports,OFRoutes} centered in bins $47.3$ miles wide. The solid line is an 8th-order polynomial fit of the marker data truncated at the boundaries.} }
\end{figure}

Another example with practical significance can be found in the analysis of airline routes in the United States.  For this case, we have analyzed all direct routes between primary airports -- i.e., airports listed by the Federal Aviation Administration (FAA) as those that provide scheduled passenger services and which process over $10,\!000$ boardings per year -- in the conterminous states.  This analysis focused on $336$ airports and $2,\!422$ unique routes.  Data was obtained from~\cite{OFAirports,OFRoutes}.  A histogram of the connection probability, viewed as a function of distance between cities, was constructed, and a polynomial fit was produced from this data.  \clr{An 8th order polynomial was chosen to ensure an adequate goodness of fit was achieved (sum of squared errors: $2.36\times 10^{-3}$) while capturing the physical characteristics at the boundaries, i.e., the probability of connectivity should go to zero for short and long node separations.  The polynomial was truncated at these boundaries.}  The average connection probability and correlation between distance and the connection function were calculated to be $k_1 = 0.043$ and $k_2 = 28.990$ miles, respectively.  These constraints were then applied \clr{separately} to compute the \clr{corresponding} maximum entropy connection functions.  \clr{The maximum entropy function was also computed when both constraints are active.}  The polynomial fit and the maximum entropy function are depicted in Fig.~\ref{fig:flights}.  There is clearly a discrepancy between \clr{each of the maximum entropy curves and the interpolated observed connection function}.  Yet it is interesting to note that the entropy resulting from the observed connection function is greater than $99.5\%$ that of the maximum predicted value \clr{when both constraints are applied}.  \clr{However, this high value should be benchmarked.  Comparing this to the case where only the mean constraint is applied, we found that the ratio was $95.3\%$, a high value despite the dissimilar connection functions (Fig.~\ref{fig:flights}).  When only the correlation constraint is applied, the ratio is $84.9\%$.}

It is important to note that the idea of ensemble entropy viewed in this context relates to a \emph{typical} flight network derived from an identical connection rule to that observed in this example.  To elucidate this point, observe that the pair connection function estimated here is taken from data that describes a single network.  So the Shannon entropy of the ensemble associated with this connection function would appear to bear little value.  Yet, we know that the entropy of the ensemble is approximately equal to the logarithm of the size of the typical set of networks, and any particular network belonging to the typical set is observed with probability approximately equal to the reciprocal of the size of the set.  The US flight network can be viewed as a member of the typical set of networks that arise from the corresponding pair connection function and pair distance distribution.  Through this reasoning, we can relate the notion of entropy defined in eq.~(\ref{eq:H(G)_1}) to our study of a single realization of a practical network.

\begin{figure}
\includegraphics[width=8cm]{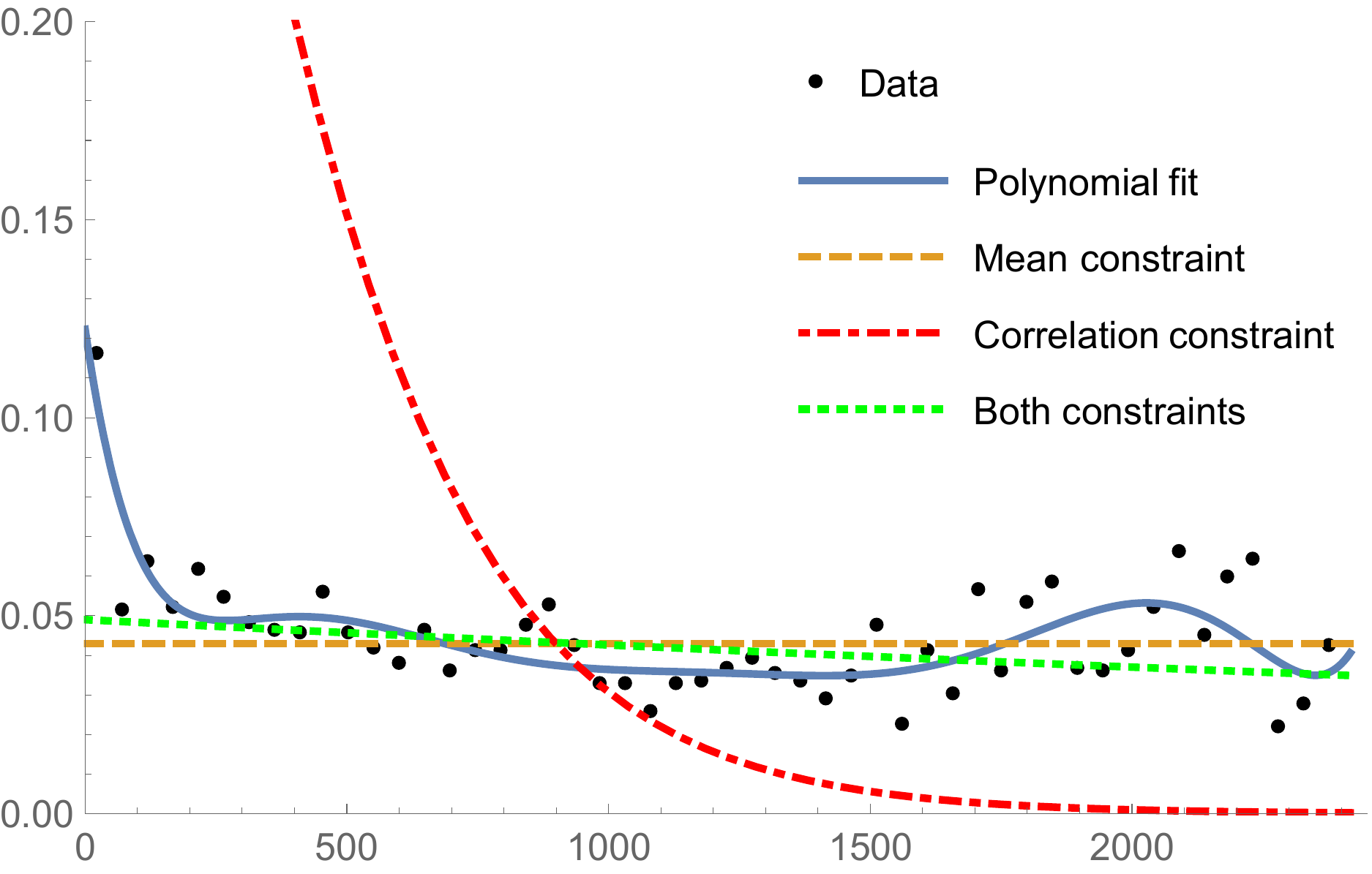}%
\caption{\label{fig:randflights}\clr{Pair connection function derived from US flight data by executing a degree-preserving randomization procedure (solid line) and maximum conditional entropy functions for different sets of constraints (dashed lines) plotted as a function of the distance between primary airports in the conterminous US.  All parameters are the same as for Fig.~\ref{fig:flights}.} }
\end{figure}

\clr{To explore this example further, degree-preserving edge permutations were carried out on the airline route data~\cite{Maslov2002,MaslovCode}, and the randomized networks were analyzed in a similar manner.  All examples that were studied exhibited very similar behavior.  The results for one such study are shown in Fig.~\ref{fig:randflights}.  Predictably, the randomization procedure had the effect of ``softening'' the pair connection function, since the permutations were made independent of internode distance.  Interestingly, the entropy resulting from the observed connection function in the randomized network is greater than $99.6\%$ of the maximum predicted value when both constraints are applied.  The ratio when only the mean constraint is applied is also approximately $99.6\%$, and the ratio corresponding to just the active correlation constraint is $73.7\%$.  Although these results may first appear to contradict those obtained for the original data, it should be noted that the maximum entropy function in this example is matched to the new constraints, and hence the observations are reasonable.  More importantly, the conclusion that we can draw from these two examples and that of the synthetically generated wireless network is that by limiting the network through both mean and correlation constraints matched to practical systems, we observe nearly maximally entropic behavior.  
}

\section{\label{sec:con}Conclusions}
In this paper, we introduced a comprehensive framework for analyzing the Shannon entropy of spatial network ensembles.  We first derived a simple upper bound on the (unconditional) entropy of an ensemble.  The tightness of the bound relates to the degree of independence observed in the set of edge states.  We then defined the  conditional entropy of a spatial network ensemble to be the entropy of the graph topology ensemble conditioned on the underlying statistical distribution of the node locations.  This formalism was exploited to calculate entropy maximizing pair connection functions conditioned on various constraints.  Applying this result to two physical examples illustrated that maximally entropic engineered systems exist in practice.

Numerous extensions and modifications of the framework detailed herein can be explored.  For example, characterizing the entropy of temporal network ensembles is a natural task that could shed light on spatial network dynamics.  Directed network ensembles can also be studied.  Furthermore, one could define different connection functions for different layers in a spatial multiplex and apply the method proposed here to study such ensembles.  In general, the formalism adopted in this work can be used to derive a general theory of information related to spatial network structure.  Along with such a theory would come operational interpretations of entropy in spatial networks -- e.g., minimum description length of the network topology -- and these interpretations will lead to better understanding and more optimal design (where applicable) of physical networks.  

\newpage
\begin{acknowledgments}
This work was supported by EPSRC grant numbers EP/N002350/1 and EP/N002458/1.
\end{acknowledgments}


%

\end{document}